\documentstyle[epsfig,12pt]{article}

\begin{document}

\author{THE STATIC CYLINDER, GYROSCOPES \and THE C-METRIC, AND ALL THAT.\bigskip
\and L. Herrera\thanks{%
Also at Deapartamento de Fisica, Facultad de Ciencias,UCV, Caracas,
Venezuela; e-mail: lherrera@gugu.usal.es} , J. Ruifern\'{a}ndez \\
{\small Area de F\'\i sica Te\'orica, Facultad de Ciencias,}\\
{\small Universidad de Salamanca, 37008 Salamanca, Spain.\smallskip }\\
and \and N. O. Santos\thanks{%
Also at Universit\'{e} Paris VI, Laboratoire du Gravitation et Cosmologie
Relativistes, Tour 22-12, 4 \`{e}me \'{e}tage, Bo\^{i}te142, 4 place
Inssien, 75005 Paris, France; e-mail: nos@conex.com.br} \\
{\small Laborat\'orio de Astrof\'\i sica e Radioastronomia}\\
{\small Centro Regional Sul de Pesquisas Espaciais - INPE/MCT}\\
{\small Cidade Universit\'aria, 97105-900 Santa Maria RS, Brazil.}}
\title{}
\maketitle

\begin{abstract}
The physical meaning of the Levi-Civita spacetime, for some ''critical''
values of the parameter $\sigma $, is discussed in the light of gedanken
experiments performed with gyroscopes circumventing the axis of symmetry.
The fact that $\sigma =1/2$ corresponds to flat space described from the
point of view of an accelerated frame of reference, led us to incorporate
the C-metric into discussion. The interpretation of $\phi $ as an angle
coordinate for any value of $\sigma $, appears to be at the origin of
difficulties.\bigskip
\end{abstract}

\newpage

\section{Introduction}

To provide physical meaning to solutions of Einstein equations, is an
endeavour whose relevance deserves to be emphasized \cite{1}.

This is particularly true in the case of the Levi-Civita (LC) spacetime \cite
{2} which after many years and a long list of works dedicated to its
discussion still presents serious challenges to its interpretation (\cite{1}%
, \cite{3}-\cite{14}, and references therein).

This metric has two essential constants, usually denoted by $a$ and $\sigma $%
. One of them, $a,$ has to do with the topology of spacetime and, more
specifically, refers to the deficit angle. It may accordingly be related to
the gravitational analog of Aharonov Bohmm effect \cite{15}, \cite{16}.

It is however $\sigma $, the parameter which presents the most serious
obstacles to its interpretation.

Indeed, for small $\sigma $ $(\sigma \leq 1/4),$ LC describes the spacetime
generated by an infinite line mass, with mass $\sigma $ per unit coordinate
length. When $\sigma =0$ the spacetime is flat \cite{1}.

However, circular timelike geodesics exits only for 
\begin{equation}
1/4>\sigma >0,  \label{1}
\end{equation}
becoming null when $\sigma =1/4$ and being spacelike for $\sigma >1/4.$

Furthermore, as the value of $\sigma $ increases from $1/4$ to $1/2$ the
corresponding Kretschmann scalar diminishes monotonically, vanishing at $%
\sigma =1/2$, and implying thereby that the space is flat also when $\sigma
=1/2.$

Still worse, if $\sigma =-1/2$ the spacetime admits an extra Killing vector
which corresponds to plane symmetry \cite{7} (also present of course in the $%
\sigma =1/2$ case).

Thus, the obvious question is: What does LC represents for values of $\sigma 
$ outside the range (0,$1/4$)?

The absence of circular test particle orbits for $\sigma >1/4$, and the fact
that most of the known material sources for LC, \cite{4}, \cite{5}, \cite{6}%
, \cite{11} require $\sigma \leq 1/4$, led to think that LC describes the
field of a cylinder only if $\sigma $ ranges within the (0,$1/4$) interval.

However, interior solutions matching to LC exist,\cite{9}, \cite{12},\cite
{14}, \cite{17} with $\sigma >1/4$.

Furthermore, the absence of circular test particle orbits for $\sigma >1/4$
may simply be interpreted, as due to the fact that the centrifugal force
required to balance the gravitational attraction implies velocities of the
test particle larger than 1 (speed of light) \cite{4}.

This last argument in turn, was objected in the past on the basis that
Kretschmann scalar decreases as $\sigma $ increases from $1/4$ to $1/2$,
suggesting thereby that the gravitational field becomes weaker \cite{9}, 
\cite{11}. However, as it has been recently emphasized \cite{12}, \cite{18},
Kretschmann scalar may not be a good measure of the strength of the
gravitational field. Instead, those authors suggest that the acceleration of
the test particle represents more suitably the intensity of the field.
Parenthetically, this acceleration increases with $\sigma $ in the interval (%
$1/4$,$1/2$).

On the basis of the arguments above and from the study of a specific
interior solution matched to LC \cite{17}, Bonnor \cite{18} proposes to
interpret LC as the spacetime generated by a cylinder whose radius increases
with $\sigma $, and tends to infinity as $\sigma $ approaches $1/2$. This
last fact suggests that when $\sigma =1/2$, the cylinder becomes a plane.
This interpretation of the $\sigma =1/2$ case was already put forward by
Gautreau and Hoffman in \cite{7} (observe that theirs $\sigma $ is twice
ours), though based on different considerations.

However, in our opinion, the question is not yet solved.

Indeed, the interior solution analyzed in \cite{18} is not valid when $%
\sigma =1/2$.

Therefore the vanishing of the normal curvatures of the coordinate lines on
the bounding surface when $\sigma \rightarrow 1/2$, suggests but does not
prove that the exterior solution with $\sigma =1/2$ has a plane source.

The LC spacetime has no horizons. According to our present knowledge of the
formation of black holes, this seems to indicate that there is an upper
limit to the mass per unit length of the line sources, and this limit has to
be below the critical linear mass, above which horizons are expected to be
formed \cite{6}.

The anisotropic fluid \cite{14} with $\sigma \leq 1$ matched to LC, produces
an effective mass per unit length that has maximum at $\sigma =1/2,$which
might explain the inexistence of horizons. Furthermore, this fact might
support too the previous acceleration representation of the field intensity.
It agrees with the result that the tangential speed $W$ of a test particle 
\cite{11} in a circular geodesics increases with $\sigma $, attaining $%
W\rightarrow \infty $ for $\sigma \rightarrow 1/2.$ The source studied in 
\cite{14} remains cylindrical for $\sigma =1/2,$ producing a cosmic string
with finite radius. However, the effective mass density by increasing up to $%
\sigma =1/2,$ and then decreasing for bigger values of $\sigma $, raises a
disturbing situation of a cylindrical distribution mass not curving
spacetime exactly at its maximum value.

On the other hand, there exists a puzzling asymmetry between the negative
and the positive mass case, for the plane source.

The point is that, as mentioned before, the $\sigma =-1/2$ case posseses
plane symmetry and furthermore test particles are repelled by the
singularity. Therefore LC with $\sigma =-1/2$, has been interpreted as the
gravitational field produced by an infinite sheet of negative mass density 
\cite{7} (though there are discrepancies on this point \cite{1}). However in
this case ($\sigma =-1/2$) the space is not flat, unlike the $\sigma =1/2$
case.

In other words, if we accept both interpretations, i.e. $\sigma =1/2$ $%
(-1/2) $ represents the field produced by an infinite plane with positive
(negative) mass density, then we have to cope with the strange fact that the
negative mass plane curves the spacetime, whereas the positive mass plane
does not. This asymmetry is, intuitively, difficult to understand.

In favor of the plane interpretation for the $\sigma =1/2$ case, point the
arguments presented in \cite{18}, although as already mentioned, they are
not conclusive.

Furthermore, even if we admit the arguments based on the principle of
equivalence, leading to the plane interpretation of the $\sigma =1/2$ case,
there is a problem with the localization of the source itself (the plane).

Indeed, it seems reasonable to assume, according to the equivalence
principle, that the physical components of curvature tensor of an
homogeneous static field, vanish everywhere, except on the source (the
plane), where they should be singular. However, when $\sigma =1/2$ the space
is flat everywhere (everywhere meaning the region covered by the patch of
coordinates under consideration), and therefore a pertinent question is:
Where is the source?

In the $\sigma =-1/2$ case, the plane interpretation is supported by the
plane symmetry of the spacetime, although objections to this interpretation
have been raised, on the basis that the proper distance between neighbouring
paths of test particles changes with time \cite{1}. However, see a comment
on this point, below eq.(10).

Also, in this case, the physical components of the curvature tensor, and the
Cartan scalars, are singular at $r=0$, revealing the existence of a source,
however they do not vanish ( except in the limit $r\rightarrow \infty $ )
and therefore the pertinent question here is: Why does the arguments based
on the equivalence principle, mentioned above, do not apply, if $\sigma
=-1/2 $ corresponds to a plane?

So, unless additional arguments are presented, we are inclined to think that
either of the interpretations (or both) are wrong.

In order to delve deeper into these questions, and with the purpose of
bringing forward new arguments, we propose here to analyze some gedanken
experiments with a gyroscope circumventing the axis of symmetry.

The obtained expression for the total precession per revolution, $\Delta
\phi $, will depend on $\sigma $. Then, analyzying the behaviour of $\Delta
\phi $ as function of different physical variables we shall be able to
provide additional elements for the interpretation of LC.

In relation to this, we shall consider also the C-metric (see\cite{19}-\cite
{21} and references therein), which, as it is well known, describes, in the
limit of vanishing mass parameter, the flat space as seen by an accelerated
observer (as the $\sigma =1/2$ case) .

As it will be seen below, the discussion presented here does not lead to
conclusive answers to the raised issues, but provides hints reinforcing some
already given interpretations and, in some cases, creating doubts abouts
formerly accepted points of view. In particular it appears that the
interpretation of the coordinate $\phi $ as an angle coordinate seems to be
untenable in some cases. A fact already brought out in \cite{7}.

At any rate, it is our hope that the results and arguments here presented,
will stimulate further discussions on this interesting problem.

The paper is organized as follows. In the next section we describe the LC
spacetime and the C-metric. In section 3 we give the expression for the
total precession per revolution of a gyroscope circumventing the axis of
symmetry and display figures indicating its dependence upon different
variables. Finally, results are discussed in the last section.

\section{Notation, conventions and the space time.}

We shall first describe the LC line element, together with the notation and
conventions used here. Next we shall briefly describe the C-metric.

\subsection{The Levi-Civita metric.}

The LC metric can be written as \cite{2}, \cite{22} 
\begin{equation}
ds^2=-ar^{4\sigma }dt^2+r^{8\sigma ^2-4\sigma }(dr^2+dz^2)+\frac{%
r^{2(1-2\sigma )}}ad\phi ^2,  \label{2}
\end{equation}
where $a$ and $\sigma $ are constants.\smallskip

The coordinates are numbered 
\begin{equation}
x^0=t,\qquad x^1=r,\qquad x^2=z,\qquad x^3=\phi 
,  \label{3}
\end{equation}
and their range are 
\begin{equation}
-\infty <t<\infty \;,\quad 0\leq r<\infty \;,\quad -\infty <z<\infty
\;,\quad 0\leq \phi \leq 2\pi ,  \label{4}
\end{equation}
with the hypersurface $\phi =0$ and $\phi =2\pi $ being identified.\smallskip

As stressed in \cite{1} neither $a$ nor $\sigma $ can be removed by
coordinate transformations, and therefore they have to be considered as
essential parameters of the LC metric.

As mentioned before, $a$ has to do with the topology of spacetime, giving
rise to an angular deficit $\delta $ equal to \cite{15} 
\begin{equation}
\delta =2\pi \left( 1-\frac 1{\sqrt{a}}\right) .  \label{5}
\end{equation}

Also, as commented in the introduction, the spacetime becomes flat if $%
\sigma $ is $0$ or $1/2$.

In the first case, $\sigma =0,$ the line element (2), adopts the usual form
of the Minkoswski interval in cylindrical coordinates (except for the
presence of $a$)

In the second case, $\sigma =1/2$, the line element becomes 
\begin{equation}
ds^2=-ar^2dt^2+dr^2+dz^2+\frac{d\phi ^2}a,  \label{6}
\end{equation}
this last expression corresponding to the flat spacetime described by an
uniformly accelerated observer with a topological defect associated to $a$.

Indeed, putting $a=1$ for simplificity, the transformation 
\begin{equation}
\overline{t}=r\sinh t\;,\;\overline{x}=r\cosh t\;,\;\overline{y}=\phi \;,\;%
\overline{z}=z,  \label{7}
\end{equation}
casts (6) into 
\begin{equation}
ds^2=-d\overline{t}^2+d\overline{x}^2+d\overline{y}^2+d\overline{z}^2   \label{8}
\end{equation}

Then, the components of the four-acceleration of a particle at rest in the
frame of (6) ($r=r_0=constant$, $z=constant,$ $\phi =constant$) as measured
by an observer at rest in the Minkowski frame of (8) are

\begin{equation}
a^\mu =\frac 1{r_0}\left( \sinh t,\cosh t,0,0\right) ,  \label{9}
\end{equation}
and therefore 
\begin{equation}
\alpha =\sqrt{a^\mu a_\mu }=\frac 1{r_0}  \label{10}
\end{equation}

Thereby indicating that such a particle is accelerated, with proper
acceleration $1/r_0$. It is perhaps worth noticing that due to (4) and (7),
the range of the Minkowski coordinate $\overline{y}$ is rather unusual. Also
observe that bodies located at different points, undergo different
accelerations. This implies in turn that two bodies undergoing the same
proper acceleration do not maintain the same proper distance (see p.176 in 
\cite{23} for details).

\subsection{The C-metric.}

This metric was discovered by Levi-Civita \cite{24}, and rediscovered since
then by many authors (see a detailed account in \cite{19}).

It may be written in the form 
\begin{equation}
ds^2=A^{-2}(x+y)^{-2}\left( F^{-1}dy^2+G^{-1}dx^2+Gdz^2-Fd\tau ^2\right) 
,  \label{11}
\end{equation}
with 
\begin{equation}
F=-1+y^2-2mAy^3\ , \qquad G=1-x^2-2mAx^3,  \label{12}
\end{equation}
where $m$ and $A$ are the two constant parameters of the solution.

Introducing retarded coordinates $u$ and $R$, defined by 
\begin{equation}
Au=\tau +\int^y{F^{-1}dy},  \label{13}
\end{equation}
\begin{equation}
AR=(x+y)^{-1}\ ,  \label{14}
\end{equation}
the metric takes the form 
\begin{equation}
ds^2=-Hdu^2-2dudR-2AR^2dudx+R^2\left( G^{-1}dx^2+Gdz^2\right) ,
\label{15}
\end{equation}
with 
\[
H=-A^2R^2G\ (x-A^{-1}R^{-1}). 
\]

If $A=0$, and $m\neq 0$, the C-metric becomes Schwarzschild. But, if $m=0$
and $A\neq 0,$ then (15) may be written, with $z=\phi ,$ $x=\cos \theta $ 
\newline
as 
\[
ds^2=-\left( 1-2AR\cos \theta -A^2R^2\sin ^2\theta \right) du^2-2dudR+ 
\]
\begin{equation}
+2AR^2dud\theta \sin \theta +R^2\left( d\theta ^2+\sin ^2\theta d\phi
^2\right) ,  \label{16}
\end{equation}
which can be casted into the Minkowski line element by 
\begin{equation}
\overline{t}=\left( A^{-1}-R\cos \theta \right) \sinh Au+R\cosh Au,
\label{17}
\end{equation}
\begin{equation}
\overline{z}=\left( A^{-1}-R\cos \theta \right) \cosh Au+R\sinh Au,
\label{18}
\end{equation}
\begin{equation}
\overline{x}=R\sin \theta \cos \phi ,  \label{19}
\end{equation}
\begin{equation}
\overline{y}=R\sin \theta \sin \phi .  \label{20}
\end{equation}

Now, for a particle at rest in the ($u,R,\theta ,\phi $) frame ($%
R=R_0=constant$, $\theta =\theta _0=constant,$ $\phi =constant$) the
components of the four-acceleration as measured by an observer at rest in
the ($\overline{t}$, $\overline{x}$, $\overline{y}$, $\overline{z}$) frame,
are 
\[
a^\mu =\frac A{\left( 1-AR_0\cos \theta _0\right) ^2-A^2R_0^2}\left\{ \left(
1-AR_0\cos \theta _0\right) \sinh Au+AR_0\cosh Au,\right. 
\]
\begin{equation}
\left. ,\ 0,\ 0,\ \left( 1-AR_0\cos \theta _0\right) \cosh
Au+AR_0\sinh Au\right\} .  \label{21}
\end{equation}

Then, the absolute value of the four acceleration vector for such particle
is 
\begin{equation}
\alpha =\sqrt{a^\mu a_\mu }=\frac A{\sqrt{1-2AR_0\cos \theta _0-A^2R_0^2\sin
^2\theta _0}}\smallskip  \label{22}
\end{equation}
indicating that the locus $R_0=0$ is accelerated with constant proper
acceleration $A$. Observe that in this case the ($u,R,\theta ,\phi $)
coordinates are only restricted by 
\begin{equation}
\overline{t}+\overline{z}>0  \label{23}
\end{equation}

\section{Gyroscope Precession.}

\subsection{Precession in the Levi-Civita metric.}

Let us consider a gyroscope circumventing the symmetry axis along a circular
path (not a geodesic), with angular velocity $\omega $. Then it can be shown
that the total precession per revolution is given by (see \cite{25} for
details) 
\begin{equation}
\Delta \phi =2\pi \left( 1-\frac{n\sqrt{a}\;r^{-(1-n)^2/4}}{(a^2-\omega
^2r^{2n})^{1/2}}\right) ,  \label{24}
\end{equation}
with $n=1-4\sigma $.

The tangential velocity of particles along circular trayectories (not
necessarily geodesics) on the plane ortogonal to the symmetry axis, is given
by the modulos of the four-vector (see \cite{26}, \cite{27},\cite{28}) 
\begin{equation}
W^\mu =\left[ (-g_{00})^{1/2}\left( dx^0+\frac{g_{0i}}{g_{00}}dx^i\right)
\right] ^{-1}V^\mu ,  \label{25}
\end{equation}
with 
\begin{equation}
V^\mu =(0,0,0,d\phi ).  \label{26}
\end{equation}

Then, for a particle in LC spacetime 
\begin{equation}
W=(W^\mu W_\mu )^{1/2}=\frac{r^n}a\;\omega .  \label{27}
\end{equation}
In terms of $W$, the expression for $\Delta \phi $ becomes 
\begin{equation}
\Delta \phi =2\pi \left[ 1-\frac{n\;r^{-(1-n)^2/4}}{\sqrt{a}\;\left(
1-W^2\right) ^{1/2}}\right] .  \label{28}
\end{equation}

\subsection{Precession in the C-metric.}

Next, due to the similarity of interpretation, mentioned before, between the 
$\sigma =1/2$ case and the C-metric with $m=0$, we shall also calculate the
total precession per revolution of a gyroscope circumventing the axis of
symmetry, in the space-time of the C-metric.

Using the Rindler-Perlick method \cite{29}, and writing the C-metric in the
form \cite{20} 
\[
ds^2=-Hdt^2+\frac{dR^2}H-\frac{2\sin {\theta }\,\cos {\theta }}{H\;p(1+3Amp)}%
AR^2dR\,d\theta + 
\]
\begin{equation}
+\frac{R^2\cos ^2\theta }{p^2(1+3Amp)^2}\left( 1+\frac{A^2R^2\sin ^2\theta }%
H\right) d\theta ^2+R^2\sin ^2\theta d\phi ^2\smallskip ,  \label{29}
\end{equation}
with 
\[
H=1-2ARp-A^2R^2(1-p^2)-\frac{2m}R(1-ARp)^3= 
\]
\begin{equation}
=(1-ARp)^2-A^2R^2-\frac{2m}R(1-ARp)^3,
\end{equation}
and 
\begin{equation}
\sin ^2\theta =1-p^2-2Amp^3,  \label{31}
\end{equation}
one obtains, 
\[
\Delta \phi =2\pi \left\{ 1-\left( \sin ^2\theta (1-ARp)^2\frac{\beta ^2}{R^2%
}+Hp^2(1+3Amp)^2\right) ^{1/2}\cdot \right. 
\]
\begin{equation}
\cdot \left. \left( H-\omega ^2R^2\sin ^2\theta \right) ^{-1/2}\right\}
\label{32}
\end{equation}
with 
\[
\beta =R-3m(1-Arp). 
\]

If $m=A=0$ we recover the usual Thomas precession in a Minkowski spacetime.

If $m=0$ and $A\neq 0,$ on the $\theta =\frac \pi 2$ plane, 
\begin{equation}
\Delta \phi =2\pi \left\{ 1-\left[ 1-R^2(A^2+\omega ^2)\right]
^{-1/2}\right\}  \label{33}
\end{equation}
which is the Thomas precession modified by the acceleration factor $A$;
while if $m\neq 0$, $A=0$, we recover the usual Fokker-de Sitter expression
for precession of a gyroscope in the Schwarzschild metric \cite{29}, 
\begin{equation}
\Delta \phi =2\pi \left\{ 1-\left( 1-\frac{3m}R\right) \left( 1-\frac{2m}%
R-\omega ^2R^2\right) ^{-1/2}\right\}  \label{34}
\end{equation}

In the general case $m\neq 0$, $A\neq 0$ (on the $\theta =\frac \pi 2$
plane), we have from (31) that, either $p=0$ or $p=-1/2$. In the first case (%
$p=0$) we obtain 
\begin{equation}
\Delta \phi =2\pi \left\{ 1-\left( 1-\frac{3m}R\right) \left[ 1-\frac{2m}%
R-\left( A^2+\omega ^2\right) R^2\right] ^{-1/2}\right\}  \label{35}
\end{equation}
whereas in the case $p=-1/2$ , the result is 
\[
\Delta \phi =2\pi \left\{ 1+\left( \left( \frac{3m}R+2\right) ^2+\left(
\frac Rm+\frac 32\right) -\frac{\left( R+2m\right) ^2}{32A^2m^3R}\right)
^{1/2}\cdot \right. 
\]
\begin{equation}
\left. \cdot \left[ -\frac{\left( R+2m\right) ^2}{2mR}-\left( A^2+\omega
^2\right) R^2\right] ^{-1/2}\right\}  \label{36}
\end{equation}

However, this last case implies $m<0$, for otherwise $H<0$ , what would
change the signature of the metric.

Finally, the tangential velocity of the gyroscope on the circular orbit
calculated from (23) for the C-metric yields 
\begin{equation}
W=(W^\mu W_\mu )^{1/2}=H^{-1/2}\;\omega R\sin \theta  \label{37}
\end{equation}

Then replacing $\omega $ by $W$ with (37), into (35), we obtain ($\theta
=\frac \pi 2$) 
\begin{equation}
\Delta \phi =2\pi \left\{ 1-\left( 1-\frac{3m}R\right) \cdot \left(
1-W^2\right) ^{-1/2}H^{-1/2}\right\}  \label{38}
\end{equation}
where 
\[
H=1-A^2R^2-\frac{2m}R\qquad {\rm if}\quad p=0 
\]
and 
\begin{equation}
\Delta \phi =2\pi \left\{ 1+\left( \left( \frac{3m}R+2\right) ^2+\left(
\frac Rm+\frac 32\right) -\frac{\left( R+2m\right) ^2}{32A^2m^3R}\right)
^{1/2}\left( 1-W^2\right) ^{-1/2}H^{-1/2}\right\}  \label{39}
\end{equation}
where 
\[
\rm  H=-\frac{\left( R+2m\right) ^2}{2mR}-A^2R^2\ \qquad {\rm if}%
\quad p=-\frac 1{2Am} 
\]
in the last case however, remember that $m$ must be negative.\smallskip

If $m=0$, (32) may be written ( with (37) ) as 
\begin{equation}
\Delta \phi =2\pi \left\{ 1-\frac{\sqrt{1+\alpha ^2R^2\sin ^2\theta }}{\sqrt{%
1-W^2}}\;\right\}  \label{40}
\end{equation}
with 
\begin{equation}
\alpha =(a^\mu a_\mu )^{1/2}  \label{41}
\end{equation}
indicating that the precession is retrograde for any $\alpha $ and $\theta .$

In the next section we shall discuss about the meaning of LC in the light of
the information provided by (28) and (38).

\section{Discussion}

Let us now analyze some figures obtained from (28) and (38).

Figure (1) exhibits the dependence of $\Delta \phi /2\pi $ on $n$ for
different values of $W$ (for simplicity all figures are ploted with $a=1$).
\begin{figure}[h]
\epsfig{figure=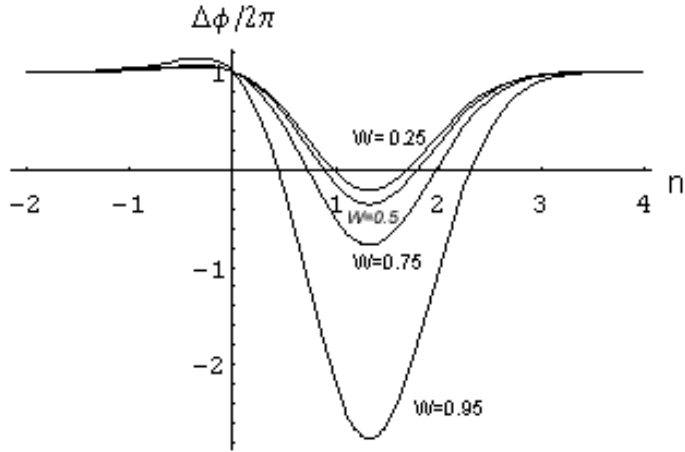,height=2.59in}
\caption{$\Delta \phi  /2 
\pi $ as function of n, for different values of W, for LC.}
\end{figure}

For $n<0$ ($\sigma >1/4$) the precession is always forward ($\Delta \phi >0$%
) as it obvious from (28). However for $n>0$ ($\sigma <1/4$) it may be
retrograde ($\Delta \phi <0$) depending on $r$ and $W$, as indicated in
figure(2), figure(3).

\begin{figure}[p]
\epsfig{figure=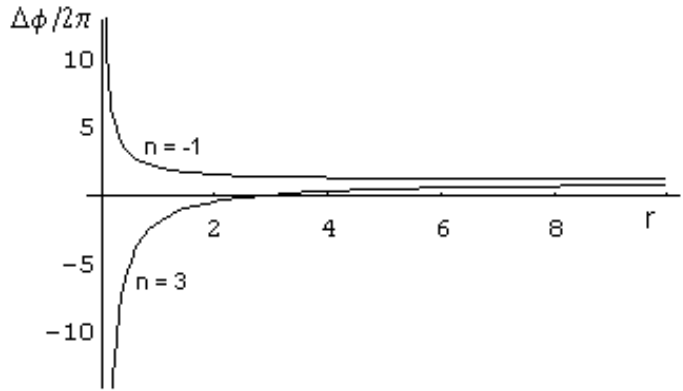,height=2.5in}
\caption{$\Delta \phi /2\pi $  as function of $r$, for $W=0.05$
 and two different values of $n$ $(-1,3)$ , for LC.}
\end{figure}

\vspace{5mm}

\begin{figure}[p]
\epsfig{figure=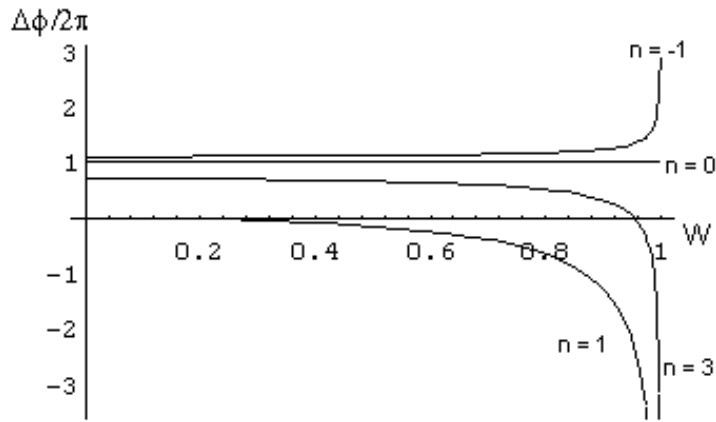,height=2.5in}
\caption{$\Delta \phi  /2
\pi$ as function of $W$, for different
values of $n$, and $r=10$, for LC.}
\end{figure}

Thus the cases $n=-1$ ($\sigma =1/2$) and $n=3$ ($\sigma =-1/2$) induce very
different behaviours on gyroscopes. This fact, together with the assymmetry
mentioned in the Introduction, reinforces our doubts about the simultaneous
interpretation of both cases ($\sigma =-1/2,1/2$) as due to infinite sheet
of either positive or negative mass density.

Next, let us consider the C-metric in the $m=0$ case. Figure(4) shows the
behaviour of the gyroscope as function of the acceleration. Observe that the
precession is retrograde, in contrast with the $n=-1$ case, for which $%
\Delta \phi $ is always positive. This behaviour is the opposite for LC and $%
n$ equal to $-1$ (see fig.(5)), and reinforces further the difficulty of
interpreting $\phi $ (in LC with $n=-1$) as the usual azhimutal angle. Still
worse, in this later case, $\Delta \phi $ always exceed $2\pi $ indicating
that the precession is forward even in the rotating frame.

\begin{figure}[h]
\epsfig{figure=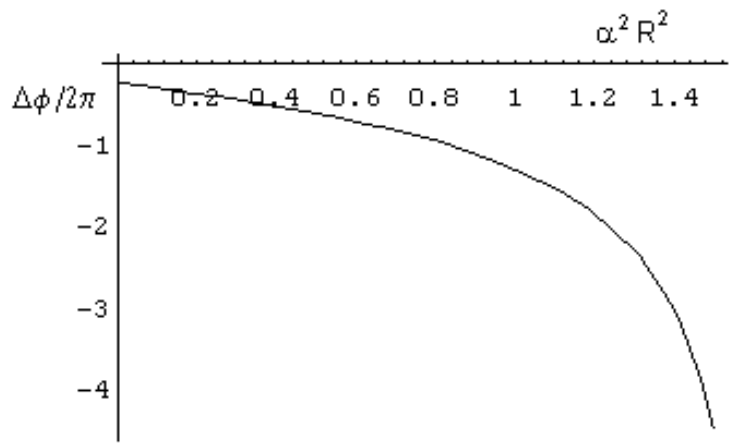,height=2.59in}
\caption{$\Delta \phi  /2
\pi $ as function of the acceleration, for the C-metric,
with $m=0$ and $\theta =\frac \pi 2$.}
\end{figure}

Now, since both cases (C-metric with $m=0$ and LC with $n=-1$) represent the
same physical situation (i.e. flat space described by an uniformly
accelerated observer) then we have to conclude that the meaning of $\phi $
in LC with $n=-1$, is different from its usual interpretation (as an angle).
This also becomes apparent from the definition of $W$ given by (27) (the
tangential velocity decreases as $1/r$). Also observe that in the case of
the C-metric with $m=0$, we recover the Thomas precession in the limit $%
\alpha =0$. This however is impossible in the LC case with $n=-1$. In the
same order of ideas it is worth noticing that in the case $n=3$, the meaning
of $\phi $ seems to correspond (qualitatively) to that of an azimuthal angle.

On the other hand, it is clear that in the case of a plane source we should
not expect $\phi $ to behave like an angle coordinate (see also \cite{12} on
this point). Therefore , on the basis of all comments above, we are inclined
to think (as in \cite{18}) that the $\sigma =$ $1/2$ case corresponds to an
infinite plane. The absence of singularities in the physical components of
the curvature tensor, remaining unexplained, although (probably) related to
the restrictions on the covering, of the coordinate system. By the same
arguments it should be clear that the interpretation of the $n=3$ case as
due to a plane, seems to be questionable.

\begin{figure}[h]
\epsfig{figure=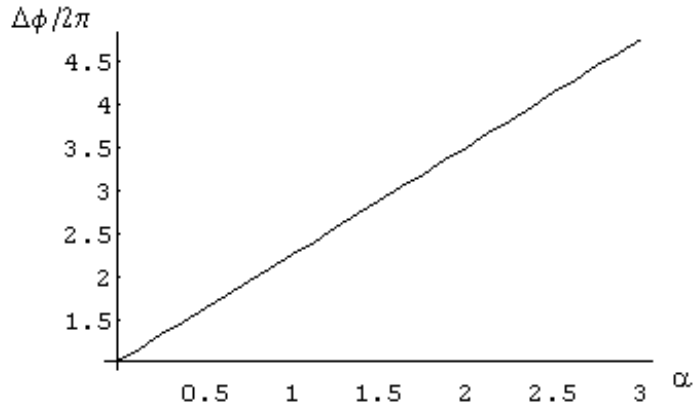,height=2.59in}
\caption{$\Delta \phi  /2 \pi $ as function
of the acceleration, for $n=-1$ in LC}
\end{figure}


\newpage

\begin{thebibliography}{99}
\bibitem{1}  W.B.Bonnor, {\it Gen.Rel.Grav.} {\bf 24}, 551 (1992).

\bibitem{2}  T.Levi-Civita, {\it Rend.Acc.Lincei} {\bf 26}, 307 (1917).

\bibitem{3}  L.Marder, {\it Proc.R.Soc. A} 244, 524 (1958)

\bibitem{4}  A.K.Raychaudhuri and M.M.Som, {\it Proc.Camb.Phil.Soc.} {\bf 58}%
, 388, (1962).

\bibitem{5}  A.F.F.Teixeira and M.M.Som, {\it Nuovo Cimento} B {\bf 21}, 64
(1974).

\bibitem{6}  J.D.Lathrop and M.S.Orsene, {\it J.Math.Phys.} {\bf 21}, 152
(1980).

\bibitem{7}  R.Gautreau and R.B.Hoffman, {\it Nuovo Cimento} B {\bf 61}, 411
(1969).

\bibitem{8}  W.B.Bonnor and M.A.P.Martins, {\it Class.Quantum Grav.} {\bf 8}%
, 727 (1991).

\bibitem{9}  W.B.Bonnor and W.Davidson, {\it Class.Quantum Grav.} {\bf 9},
2065 (1992).

\bibitem{10}  J.Stela and D.Kramer, {\it Acta Phys. Pol. B} {\bf 21}, 843
(1990).

\bibitem{11}  M.F.A.da Silva, L.Herrera, F.M.Paiva and N.O.Santos, {\it %
J.Math.Phys.} {\bf 36}, 3625 (1995).

\bibitem{12}  T.G.Philbin, {\it Class.Quantum Grav.} {\bf 13}, 1217 (1996).

\bibitem{13}  M.A.H.MacCallum, {\it Gen.Rel.Grav.} {\bf 30}, 131 (1998).

\bibitem{14}  A.Z.Wang, M.F.A.da Silva and N.O.Santos, {\it Class.Quantum
Grav.} {\bf 14}, 2417 (1997).

\bibitem{15}  J.S.Dowker, {\it Nuovo Cimento} B {\bf 52}, 129 (1967).

\bibitem{16}  M.F.A.da Silva, L.Herrera, F.M.Paiva and N.O.Santos, {\it %
Gen.Rel.Grav.} {\bf 27}, 859 (1995).

\bibitem{17}  S.Haggag and F.Desokey, {\it Class.Quantum Grav.} {\bf 13},
3221 (1996).

\bibitem{18}  W.B.Bonnor, The static cylinder in general relativity, in ''On
Einstein's Path'' ed.Alex Harvey, (New York; Springer) 1999, page 113.

\bibitem{19}  W.Kinnersley and M.Walker, {\it Phys.Rev.} D {\bf 2}, 1359
(1970).

\bibitem{20}  H.Farhoosh and and R.L.Zimmerman, {\it Phys.Rev.} D {\bf 23},
299 (1981).

\bibitem{21}  W.B.Bonnor, {\it Gen.Rel.Grav.} {\bf 15}, 535 (1983).

\bibitem{22}  D.Kramer, H.Stephani, M.A.H.MacCallum and E.Herlt, {\it Exact
Solutions of Einstein's Field Equations, (}Cambridge University Press,
Cambridge{\it ), }(1980).

\bibitem{23}  D.Boulware, {\it Ann.Phys.(N.Y)}, 124,169, (1980).

\bibitem{24}  T.Levi-Civita, {\it Atti Accad.Naz. Lincei Rend.}, {\bf 27},
343 (1918).

\bibitem{25}  L.Herrera, F.M.Paiva and N.O.Santos,
{\it Class.Quantum Grav.}, {\bf 17}, 1549 (2000).

\bibitem{26}  L.Landau and E.M.Lifshitz, {\it The Classical Theory of Fields}%
, (Reading, Massachussets) (1962).

\bibitem{27}  J.LAnderson, {\it Principles of Relativity}, (Academic Press,
New York) (1967).

\bibitem{28}  L.Herrera and N.O.Santos, {\it J.Math.Phys.} {\bf 39}, 3817
(1998).

\bibitem{29}  W.Rindler and V.Perlick, {\it Gen.Rel.Grav.} {\bf 22}, 1067
(1990).
\end{thebibliography}
\end{document}